\documentclass[fleqn,twoside]{article}
\usepackage{espcrc2}
\usepackage{graphicx}
\def\nabstar#1{\nabla\kern-0.5pt\smash{\raise 4.5pt\hbox{$\ast$}}
               \kern-4.5pt_{#1}}

\def\drvstar#1{\partial\kern-0.5pt\smash{\raise 4.5pt\hbox{$\ast$}}
               \kern-5.0pt_{#1}}

\def\newline{\relax\ifhmode\null\hfil\break\else\nonhmodeerr@\newline\fi}
\def\frac#1#2{{#1\over#2}}
\def\text#1{{\hbox{\rm #1}}}

\newcommand{\beq}{\begin{equation}}
\newcommand{\eeq}{\end{equation}}
\newcommand{\bea}{\begin{eqnarray}}
\newcommand{\eea}{\end{eqnarray}}

\def\BE{\begin{equation}}
\def\EE{\end{equation}}
\def\BA{\begin{eqnarray}}
\def\EA{\end{eqnarray}}
\def\BAN{\begin{eqnarray*}}
\def\EAN{\end{eqnarray*}}

\def\gm5{\gamma_5}

\def\BE{\begin{equation}}
\def\EE{\end{equation}}
\def\BA{\begin{eqnarray}}
\def\EA{\end{eqnarray}}
\def\BAN{\begin{eqnarray*}}
\def\EAN{\end{eqnarray*}}

\def\text#1{{\rm #1}}
\newcommand{\AmS}{{\protect\the\textfont2
  A\kern-.1667em\lower.5ex\hbox{M}\kern-.125emS}}
\title{
       \vspace{-4.0cm}
       \rightline{\normalsize NTUTH-02-505E}
       \vspace{-0.2cm}
       \rightline{\normalsize August 2002}
       \vspace{2.9cm}
Quenched chiral logarithms in lattice QCD with
overlap Dirac quarks\thanks{Talk presented by T.W. Chiu at Lattice 2002.}
\thanks{This work was supported in part by National Science Council,
        R.O.C. under the grant numbers NSC90-2112-M002-021}}

\author{Ting-Wai Chiu\address[PNTU]{Physics Department, National Taiwan
        University, Taipei, Taiwan 106, Taiwan.},
        Tung-Han Hsieh\addressmark[PNTU]
        }
       
\begin{document}

\begin{abstract}

We examine quenched chiral logarithms in lattice QCD with overlap
Dirac quarks. From our data of $ m_\pi^2 $, we determine
the coefficient of quenched chiral logarithm
$ \delta $ = 0.203(14), 0.176(17), 0.193(17) and 0.200(13)
for lattices of sizes $ 8^3 \times 24 $, $ 10^3 \times 24 $,
$ 12^3 \times 24 $ and $ 16^3 \times 32 $ respectively.
Also, for the first three lattice sizes, we measure the index
susceptibility of the overlap Dirac operator, and use the exact relation
between the index susceptibility and the $ \eta' $ mass in quenched
chiral perturbation theory to obtain an independent determination of
$ \delta = 0.198(27), 0.173(24), 0.169(22) $,
which are in good agreement with those determined from $ m_\pi^2 $.

\vspace{1pc}
\end{abstract}

\maketitle

\begin{table*}[htb]
\caption{
Determination of the coefficient of quenched chiral logarithm $ \delta $
by fitting (\ref{eq:mpi2}) to data of $ m_\pi^2 $. The $ \eta' $ mass
in q$\chi$PT is computed from (\ref{eq:delta}).
}
\begin{center}
\begin{tabular}{cccccccccc}
\hline
& & \multicolumn{2}{c}{$m_{\pi}$(MeV)}  &  &  \\
$ L^3 \times T $ & $a$(fm)
                 & min. & max. & $L/\lambda^{max}$
                 & $a/\lambda^{min}$
                 & $ C a $ & $ B $ & $ \delta $ & $m_{\eta'}$(MeV)  \\
\hline
\hline
$ 8^3 \times 24$ & 0.147(1)
                 & 418  & 1047 & 2.5 & 0.78
                 & 1.193(18) & 1.152(56)  & 0.203(14)  & 916(32) \\
$10^3 \times 24$ & 0.152(1)
                 & 375  & 834  & 2.9 & 0.64
                 & 1.017(25) & 1.222(71)  & 0.176(17)  & 853(42) \\
$12^3 \times 24$ & 0.147(1)
                 & 395  & 870  & 3.5 & 0.65
                 & 1.051(24) & 1.254(73)  & 0.193(17)  & 893(40) \\
$16^3 \times 32$ & 0.095(1)
                 & 439  & 1101 & 3.4 & 0.53
                 & 1.117(15) & 2.146(102) & 0.200(13)  & 909(31) \\
\hline
\end{tabular}
\label{tab:mpi2}
\vspace{-0.6cm}
\end{center}
\end{table*}

\section{Introduction}
In quenched chiral perturbation theory (q$\chi$PT) \cite{Sharpe:1992ft},
the pion mass to one-loop order reads as
\bea
\label{eq:mpi2}
m_\pi^2 = C m_q \{ 1 - \delta[ \mbox{ln}( C m_q/\Lambda_{\chi}^2 ) + 1 ] \}
          + B m_q^2
\eea
where $ m_q $ denotes the bare ($ u $ and $ d $) quark mass,
$ \Lambda_{\chi} $ is the chiral cutoff which can be taken to be
$ 2 \sqrt{2} \pi f_{\pi} $ ( $ f_{\pi} \simeq 132 $ MeV ),
$ C $ and $ B $ are parameters,
and $ \delta $ is the coefficient of the quenched chiral logarithm.

Theoretically, $ \delta $ can be estimated to be \cite{Sharpe:1992ft}
\bea
\label{eq:delta}
\delta = \frac{ m_{\eta'}^2 }{ 8 \pi^2 f_\pi^2 N_f }
\eea
where $ m_{\eta'} $ denotes the $ \eta' $ mass in q$\chi$PT,
and $ N_f $ is the number of light quark flavors.
For $f_{\pi}=132$ MeV, $N_f=3 $, and
$m_{\eta'}=\sqrt{{\bf m}_{\eta'}^2 + m_\eta^2 - 2 m_K^2}=853$ MeV,
(\ref{eq:delta}) gives
\bea
\label{eq:delta_cpt}
\delta \simeq 0.176
\eea
Evidently, if
one can extract $ \delta $ from the data of $ m_\pi^2 $,
then $ m_{\eta'} $ in q$\chi$PT can be determined by (\ref{eq:delta}).

Besides from the data of $ m_{\pi}^2 $,
one can also obtain $ \delta $ via (\ref{eq:delta})
by extracting $ m_{\eta'} $ from the propagator
of the disconnected hairpin diagram.
However, to compute the propagator of the hairpin is very tedious.

Fortunately, with the realization of exact chiral symmetry on the
lattice \cite{Neuberger:1998fp,Narayanan:1995gw},
the quark propagator coupling to
$ \eta' $ is $ ( D_c + m_q )^{-1} $ \cite{Chiu:1998eu},
thus only the zero modes of $ D_c $ can contribute
to the hairpin diagram, regardless of the bare quark mass $ m_q $.
Therefore one can derive an exact relation \cite{Chiu:2002xm} between
the $ \eta' $ mass in q$\chi$PT
and the index susceptibility of any Ginsparg-Wilson lattice
Dirac operator, without computing the hairpin
diagram at all. Explicitly, this exact relation reads as
\bea
\label{eq:wv}
( m_{\eta'} a )^2
     = \frac{4 N_f }{(f_{\pi}a)^2}
       \frac{\langle (n_{+}-n_{-})^2 \rangle }{N_s}
\eea
where $ N_s $ is the total number of sites, and
$ \chi \equiv \langle (n_{+} - n_{-})^2 \rangle / N_s $ is the
index susceptibility of any Ginsparg-Wilson lattice Dirac operator
in the quenched approximation.
Then (\ref{eq:delta}) and (\ref{eq:wv}) together gives
\bea
\label{eq:delta_s}
\delta
= \frac{ 1 }{ 2 {\pi}^2 ( f_{\pi} a )^4 }
  \frac{ \left< ( n_{+} - n_{-} )^2 \right> }{N_s}
\eea
A salient feature of (\ref{eq:delta_s}) is that $ \delta $
can be determined at finite lattice spacing $ a $, by measuring
the index (susceptibility) of the overlap Dirac operator, and
with $ f_{\pi} a $ extracted from the pion propagator.

Now it is clear that, in order to confirm the presence of
quenched chiral logarithm in lattice QCD,
one needs to check whether the coefficient $ \delta $
obtained by fitting (\ref{eq:mpi2}) to the data of $ m_{\pi}^2 $,
agrees with that (\ref{eq:delta_s}) from the index susceptibility.
This is a requirement for the consistency of
the theory, since the quenched chiral logarithm in $ m_{\pi}^2 $
(\ref{eq:mpi2}) is due to the $ \eta' $ loop coupling to the
pion propagator through the mass term (in the chiral lagrangian),
thus $ \delta $ ($ m_{\eta'} $) must be the same in both cases.
We regard this consistency requirement as a basic criterion for
lattice QCD (with any fermion scheme) to realize QCD chiral
dynamics in continuum.

Recently, we have determined $\delta=0.203(14)$,
from the data of $ m_\pi^2 $,
as well as $\delta=0.197(27)$ from the index susceptibility,
for the $ 8^3 \times 24 $ lattice \cite{Chiu:2002xm}.
Their excellent agreement suggests that lattice QCD with
overlap Dirac quarks indeed realizes the QCD chiral dynamics
in the continuum.

In this paper, we extend our studies to other lattices of larger
volumes and/or smaller lattice spacings.
Our computations are performed with a Linux PC cluster
of 30 nodes at the speed $\sim30$ Gflops \cite{Chiu:2002bi}.

\section{Results}
With the Wilson $ SU(3) $ gauge action and Creutz-Cabbibo-Marinari
heat bath algorithm, we generate ensembles of
gauge configurations as follows\footnote{The 56 configurations
on the $ 16^3 \times 32 $ lattice are retrieved from the gauge connection
(http://qcd.nersc.gov/).}.
\begin{center}
\begin{tabular}{ccc}
\hline
$ L^3 \times T $ & $ \beta $ &  \# configs. \\
\hline
\hline
$8^3  \times 24$  & 5.8  & 100 \\
$10^3 \times 24$  & 5.8  & 100 \\
$12^3 \times 24$  & 5.8  & 100  \\
$16^3 \times 32$  & 6.0  & 56   \\
\hline
\end{tabular}
\end{center}

For each configuration, quark propagators are computed
for 10-12 bare quark masses.
Then the pion propagator and its time correlation function $ G(t) $
are obtained, and $ \langle G(t) \rangle $ is fitted by the usual
formula
\BAN
\label{eq:Gt_fit}
G_{\pi}(t) = \frac{Z}{2 m_{\pi} a }
             [ e^{-m_{\pi} a t} + e^{-m_{\pi} a (T-t)} ]
\EAN
to extract the pion mass $ m_{\pi} a $ and the pion decay constant
\BAN
\label{eq:fpi}
f_{\pi} a = 2 m_q a \frac{\sqrt{Z}}{m_{\pi}^2 a^2 }
\EAN
The data of $ f_{\pi} a $ can be fitted by straight line.
Thus taking $ f_\pi a $ at $ m_q a = 0 $ equal to
$ 132 $ MeV times the lattice spacing $ a $,
then $ a $ can be determined.
Fixing $ \Lambda_{\chi} a = 2 \sqrt{2} \pi f_\pi a $,
we fit (\ref{eq:mpi2}) to the data of $ ( m_\pi a )^2 $, and obtain
$ Ca $, $ \delta $ and $ B $. Then $ m_\eta' $ is computed from
(\ref{eq:delta}). Our results are summarized in Table \ref{tab:mpi2}.
Evidently, all $ \delta $ values are in good agreement with the theoretical
estimate $ \delta \simeq 0.176 $ from q$\chi$PT.
Furthermore, the $ \eta' $ masses are also in good agreement with the
theoretical estimate $ m_{\eta'} \simeq 853 $ MeV.

\begin{figure}[htb]
\vspace{-0.6cm}
\begin{center}
\hspace{0.0cm}\includegraphics*[height=8cm,width=6.6cm]{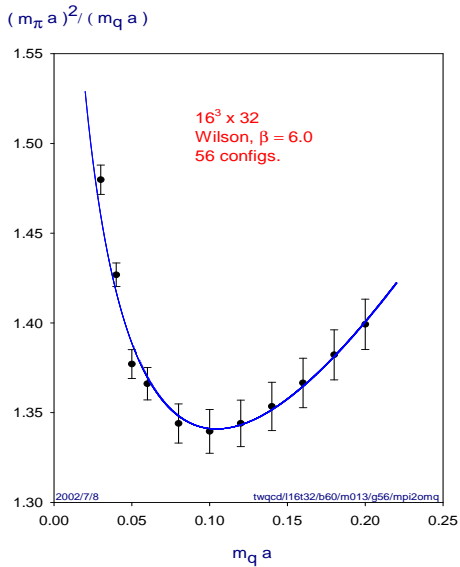}
\vspace{-1.0cm}
\caption{$(m_\pi a)^2 / (m_q a) $ versus the bare quark mass $ m_q a $.}
\label{fig:mpi2omq}
\end{center}
\vspace{-0.6cm}
\end{figure}

\begin{figure}[htb]
\vspace{0.0cm}
\begin{center}
\hspace{0.0cm}\includegraphics*[height=8cm,width=6.6cm]{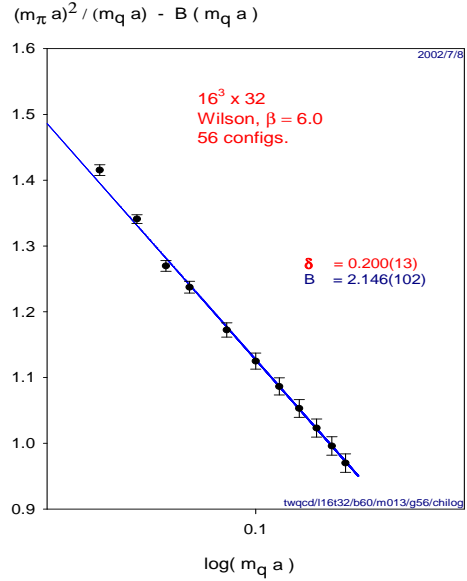}
\vspace{-1.0cm}
\caption{The extraction of the quenched chiral logarithm by plotting
$(m_\pi a)^2 / (m_q a) - B (m_q a) $ versus $ \log(m_q a) $.}
\label{fig:chiralog}
\vspace{-0.6cm}
\end{center}
\end{figure}

Even though the quenched chiral logarithm may not be easily
detected in the graph of $ (m_\pi a)^2 $ vs. $ m_q a $,
it can be unveiled by plotting $ (m_\pi a)^2 / (m_q a) $ vs. $ m_q a $,
as shown in Fig. \ref{fig:mpi2omq}.
Further, plotting $ (m_\pi a)^2 / (m_q a) - B (m_q a) $
vs. $ \log(m_q a) $, the presence of quenched chiral logarithm is evident,
as shown in Fig. \ref{fig:chiralog}.

Note that, in Fig. \ref{fig:mpi2omq},
for the $ 16^3 \times 32 $ lattice at $ \beta = 6.0 $,
the minimum of $ (m_\pi a)^2/(m_q a) $ occurs at
$ m_q a = Ca \delta / B \simeq 0.1 $, which corresponds to
$ m_q \simeq 200 $ MeV for $ a \simeq 0.095 $ fm (Table \ref{tab:mpi2}).
Similarly, for $ 8^3 \times 24 $, $ 10^3 \times 24 $,
and $ 12^3 \times 24 $ lattices at $ \beta = 5.8 $, the minima of
$ (m_\pi a)^2/(m_q a) $ occur at
$ m_q a \simeq $ 0.21, 0.15 and 0.16
(i.e., $ m_q \simeq $ 280, 190, and 217 MeV) respectively. Now,
except for the $ 8^3 \times 24 $ lattice which has the smallest volume,
the minima of $ (m_\pi a)^2/(m_q a) $ for the other three lattice sizes
almost occur at the same $ m_q \simeq 200 \pm 20 $ MeV.
This provides rather substantial evidence that the quenched chiral logarithm
parameters extracted from our data of $ m_\pi^2 $ is
indeed genuine, not due to any finite size effects or discretization errors.

Next we measure the index susceptibility of the overlap Dirac operator
by the spectral flow method \cite{Narayanan:1995gw}, and then determine
$ \delta $ via (\ref{eq:delta_s}). The $ \eta' $ mass in q$\chi$PT
is computed from (\ref{eq:wv}). Our results are summarized in
Table \ref{tab:index_s}, including our earlier results for
the $ 8^3 \times 24 $ lattice. Evidently, each
$ \delta $ is in good agreement with the corresponding $ \delta $
extracted from $ m_\pi^2 $ (Table \ref{tab:mpi2}), as well as with the
theoretic estimate $ 0.176 $.

\begin{table}
\caption{
Determination of $\delta$ (\ref{eq:delta_s}) from index susceptibility.
The $ \eta' $ mass in q$\chi$PT is computed from (\ref{eq:wv}).
}
\begin{center}
\begin{tabular}{cccc}
\hline
$ L^3 \times T $ & $ \chi/10^{-4} $ & $ \delta $ & $m_{\eta'}$(MeV) \\
\hline
\hline
$ 8^3 \times 24$ & $ 3.67(50) $ & 0.198(27)  & 904(61)  \\
$10^3 \times 24$ & $ 3.65(50) $ & 0.173(24)  & 817(58)  \\
$12^3 \times 24$ & $ 3.13(41) $ & 0.169(22)  & 835(55) \\
\hline
\end{tabular}
\end{center}
\vspace{-0.6cm}
\label{tab:index_s}
\end{table}

\section{Conclusion}
Our results in this paper provide strong evidences that
quenched QCD with overlap Dirac quarks realizes quenched
QCD chiral dynamics, as depicted by quenched chiral perturbation theory.

\end{document}